\documentclass[aps,prl,twocolumn,superscriptaddress]{revtex4}%
\usepackage{epsfig,dsfont,amssymb,amsmath,amsthm,amsfonts,amsbsy,mathrsfs}
\usepackage{graphicx}
\usepackage{epstopdf}
\usepackage{amsmath}
\usepackage{amssymb}%
\begin{document}

\title{First-principles study of two-dimensional van der Waals heterojunctions}

\author{Wei Hu}
\affiliation{Hefei National Laboratory for Physical Sciences at
Microscale, University of Science and Technology of China, Hefei,
Anhui 230026, China} \affiliation{Computational Research Division,
Lawrence Berkeley National Laboratory, Berkeley, CA 94720, USA}


\author{Jinlong Yang}
\thanks{Corresponding author. E-mail: jlyang@ustc.edu.cn}
\affiliation{Hefei National Laboratory for Physical Sciences at
Microscale, University of Science and Technology of China, Hefei,
Anhui 230026, China} \affiliation{Synergetic Innovation Center of
Quantum Information and Quantum Physics, University of Science and
Technology of China, Hefei, Anhui 230026, China}

\date{\today}

\pacs{ }

\begin{abstract}

Research on graphene and other two-dimensional (2D) materials, such
as silicene, germanene, phosphorene, hexagonal boron nitride (h-BN),
graphitic carbon nitride (g-C$_3$N$_4$), graphitic zinc oxide
(g-ZnO) and molybdenum disulphide (MoS$_2$), has recently received
considerable interest owing to their outstanding optoelectronic
properties and wide applications. Looking beyond this field,
combining the electronic structures of 2D materials in ultrathin van
der Waals heterojunctions has also emerged to widely study
theoretically and experimentally to explore some new properties and
potential applications beyond their single components. Here, this
article reviews our recent theoretical studies on the structural,
electronic, electrical and optical properties of 2D van der Waals
heterojunctions using density functional theory calculations,
including the Graphene/Silicene, Graphene/Phosphorene,
Graphene/g-ZnO, Graphene/MoS$_2$ and g-C$_3$N$_4$/MoS$_2$
heterojunctions. Our theoretical simulations, designs and
calculations show that novel 2D van der Waals heterojunctions
provide a promising future for electronic, electrochemical,
photovoltaic, photoresponsive and memory devices in the experiments.

\end{abstract}




\maketitle


\section{Introduction}

Two-dimensional (2D) ultrathin materials, such as
graphene~\cite{Scinece_306_666_2004, NatureMater_6_183_2007,
RMP_81_109_2009}, silicene~\cite{PRL_102_236804_2009,
PRL_108_155501_2012, SSR_67_1_2012, PSS_90_46_2015},
germanene~\cite{ACSNano_7_4414_2013, AdvMater_26_4820_2014,
NJP_16_095002_2014}, phosphorene~\cite{NatureNanotech_9_372_2014,
ACSNano_8_4033_2014, NatureCommun_5_4475_2014}, hexagonal boron
nitride (h-BN)~\cite{NatureMater_3_404_2004, NanoLett_10_3209_2010,
NanoLett_10_4134_2010}, graphitic carbon nitride
(g-C$_3$N$_4$)~\cite{NatureMater_8_76_2009, JACS_131_1680_2009,
JPCL_3_3330_2012}, graphitic zinc oxide
(g-ZnO)~\cite{JMC_15_139_2005, PRL_96_066102_2006,
PRL_99_026102_2007} and molybdenum disulphide
(MoS$_2$)~\cite{PRL_105_136805_2010, NatureNanotechnol_6_147_2011,
ACSNano_6_74_2012}, have received considerable interest recently
owing to their outstanding properties and wide applications. There
already have been many review articles~\cite{PNAS_102_10451_2005,
Nanoscale_3_20_2011, AdvMater_24_210_2012,
NatureNanotechnol_7_699_2012, ChemRev_113_3766_2013,
ACSNano_7_2898_2013, small_11_640_2014} for the research on 2D
materials over the past several years.

Graphene~\cite{NatureMater_6_183_2007}, a 2D sp$^2$-hybridized
carbon monolayer, is known to have remarkable electronic properties,
such as a high carrier mobility, but the absence of a bandgap limits
its applications of large-off current and high on-off ratio for
graphene-based electronic devices. The same limitation also exists
in silicene~\cite{PRL_102_236804_2009} and
germanene~\cite{ACSNano_7_4414_2013}, which have most similar
remarkable electronic properties to graphene but with buckled
honeycomb structures. On the other hand, most of graphene's
derivatives, such as graphene oxide~\cite{Nature_448_457_2007_GO,
AdvMater_22_3906_2010_GO, ChemSocRev_39_228_2012_GO},
graphane~\cite{PRB_75_153401_2007_Graphane,
Nanotechnology_20_465704_2009_Graphane,
Science_323_610_2009_Graphane} and
fluorographene~\cite{small_6_2877_2010_Fluorographene,
small_7_965_2011_Fluorographene, ACSNano_5_1042_2011_Fluorographene}
are all semiconductors but with lower carrier mobility, and have
also been studied experimentally and theoretically.

Phosphorene, a new two dimensional (2D) elemental
monolayer~\cite{NatureNanotech_9_372_2014}, has recently been
experimentally isolated through mechanical exfoliation from bulk
black phosphorus. Phosphorene exhibits some remarkable electronic
properties superior to graphene. For example, phosphorene is a
direct semiconductor with a high hole
mobility~\cite{ACSNano_8_4033_2014}, showing the drain current
modulation up to 10$^5$ in
nanoelectronics~\cite{NatureCommun_5_4475_2014}. Besides graphene,
phosphorene is the only stable elemental 2D monolayer which can be
mechanically exfoliated experimentally~\cite{ACSNano_8_4033_2014}.
Most recently, arsenene and antimonene have also predicted
theoretically~\cite{Angew_54_3112_2015}. Monolayer arsenene and
antimonene are indirect wide-band-gap semiconductors, but become
direct band-gap semiconductors under strain. Owing to these band-gap
transitions, they can be used in nanoelectronic and optoelectronic
devices.

2D h-BN monolayer~\cite{NatureMater_3_404_2004}, so-called "white
graphene", is comprised of alternating boron and nitrogen atoms in a
sp$^2$-bonded 2D honeycomb arrangement and isolated from bulk BN. It
has a wide bandgap used for its eccellent electrical insulations,
high thermal conductivity and superior lubricant properties, and can
be useful as an important complementary 2D dielectric substrate for
graphene electronics.

2D g-C$_3$N$_4$ monolayer~\cite{NatureMater_8_76_2009}, which can be
made by polymerization of cyanamide, dicyandiamide or melamine, is a
semiconductor with a bandgap of about 2.7 $eV$, as a new polymeric
photocatalyst for hydrogen production from water splitting under
visible light, which has been demonstrated to be a be a potential
candidate for solar cell absorber and photovoltaic materials.

ZnO nanofilms and nanoparticles have been widely used with graphene
as composite materials for light-emitting
diodes~\cite{Science_330_655_2010},
photocatalysts~\cite{ACSNano_4_4174_2010}, quantum
dots~\cite{Nature_7_465_2012} and field effect
transistors~\cite{Nanoscale_4_3118_2012}. However, first-principles
calculations~\cite{JMC_15_139_2005, PRL_96_066102_2006} have
predicted and then confirmed
experimentally~\cite{PRL_99_026102_2007} that the ZnO(0001) film
prefers a graphitic honeycomb structure as the layer number is
reduced. Furthermore, 2D g-ZnO monolayer itself is also
semiconducting with a wide band gap of 3.57 $eV$, showing many
interesting electronic and magnetic
properties~\cite{APL_97_122503_2010, PRB_81_195413_2010,
JPCC_116_11336_2012}.

2D MoS$_2$ is an important kind of transition metal dichalcogenides
(TMDCs), which has been widely studied experimentally and
theoretically~\cite{PRL_105_136805_2010,
NatureNanotechnol_6_147_2011, ACSNano_6_74_2012}. MoS$_2$ has a
direct band gap of about 2 $eV$, showing a high on-off current ratio
with a high carrier mobility of around 200-500 cm$^2$/(Vs) in
nanotransistors. Most of other
TMDCs~\cite{NatureNanotechnol_7_699_2012}, such as MoSe$_2$, WS$_2$
and WSe$_2$, but have attracted relatively little attention despite
the fact that they have similar electronic structures to MoS$_2$.

In parallel with the great efforts on 2D materials, another research
field has recently emerged has been gaining strength over the past
several years. It deals with ultrathin 2D van der Waals
heterojunctions and devices made layer by layer from 2D materials,
stacking different 2D monolayers on the top of each other. Strong
covalent bonds provide in-plane stability of 2D monolayers, whereas
relatively weak van der Waals forces are sufficient to keep the
stack together. The possibility of making 2D van der Waals
heterojunctions has been demonstrated experimentally for electronic,
electrochemical, photovoltaic, photoresponsive and memory devices.
Recently, there only have been several review
articles~\cite{SSC_152_1275_2012, Nature_499_419_2013,
Nanoscale_6_12250_2014, PSS_90_21_2015} involved for the research on
2D van der Waals heterojunctions.

Up to now, significant experimental and theoretical efforts have
been made to characterize and explore new structures, properties and
applications of 2D van der Waals heterojunctions, especially
graphene based heterojunctions, such as graphene/hexagonal boron
nitride (G/h-BN)~\cite{PRB_76_073103_2007, Nature_5_722_2010,
APL_98_083103_2011, NatureMater_10_282_2011, NanoLett_12_714_2012},
graphene/silicene (G/S)~\cite{JCP_139_154704_2013,
PRB_88_245408_2013, APL_103_261904_2013}, graphene/phosphorene
(G/P)~\cite{JMCC_3_4756_2015, PRL_103_066803_2015,
ACSNano_9_4138_2015}, graphene/graphitic carbon nitride
(G/g-C$_3$N$_4$)~\cite{JPCC_115_7355_2011, JACS_133_8074_2011,
JACS_134_4393_2012}, graphene/graphitic zinc oxide
(G/g-ZnO)~\cite{JCP_138_124706_2013, JPCC_117_10536_2013,
JAP_113_054307_2013}, graphene/molybdenum disulphide
(G/MoS$_2$)~\cite{Nanoscale_3_3883_2011, JPCC_117_15347_2013,
Science_340_1311_2013, ACSNano_7_7021_2013, arXiv:1411.0357_2014},
graphene/molybdenum disulfide
(G/MoSe$_2$)~\cite{Science_331_568_2011, JPCC_115_20237_2011,
JMCA_2_360_2014}, silicene/hexagonal boron nitride
(S/h-BN)~\cite{NanoLett_12_113_2012, JPCC_117_10353_2013,
PRB_87_235435_2013}, silicene/molybdenum disulphide
(S/MoS$_2$)~\cite{NanoscaleResLett_9_110_2014, JPCC_118_19129_2014,
PRB_16_11673_2014}, and TMDCs based
heterojunctions~\cite{Nanoscale_6_2879_2014, Nanoscale_6_4566_2014,
JPCA_2_7960_2014}. These van der Waals heterojunctions show much
more new properties far beyond their single components. Furthermore,
most of them are ideal substrates for each other to preserve their
intrinsic electronic properties due to weak van der Waals
interactions at the interface.

Hybrid G/h-BN heterojunctions~\cite{PRB_76_073103_2007,
Nature_5_722_2010, APL_98_083103_2011, NatureMater_10_282_2011,
NanoLett_12_714_2012} are the most widely studied topic in 2D van
der Waals heterojunctions during the past few years. Theoretically,
Giovannetti $et$ $al.$~\cite{PRB_76_073103_2007} proved that
substrate-induced band gaps can be opened at the Dirac point of
graphene on h-BN surfaces and the gap values are tunable as varying
the the interfacial distance, which greatly improves the room
temperature pinch-off characteristics of graphene-based field effect
transistors. Experimentally, Dean $et$
$al.$~\cite{Nature_5_722_2010} first in 2010 introduced atomically
thin 2D van der Waals heterojunctions in which graphene is on top of
h-BN layers, and shown that h-BN can be used as an excellent
complementary 2D dielectric substrate for graphene electronics.

In the present work, we review our recent theoretical studies on the
structural, electronic, optical and electrical properties of 2D van
der Waals heterojunctions with density functional theory
calculations and shows that these 2D van der Waals heterojunctions
provide a promising future for electronic, electrochemical,
photovoltaic, photoresponsive and memory
devices~\cite{SSC_152_1275_2012, Nature_499_419_2013,
Nanoscale_6_12250_2014, PSS_90_21_2015}.



\section{Theoretical models}

In order to simulate hybrid composite structures of 2D van der Waals
heterojunctions made layer by layer, we need to choose two special
supercells for each 2D material with different crystal systems and
lattice constants, making them have a smaller lattice mismatch for
each other. We have written a small program named LatticeMatch(See
the supplementary material) to construct theoretical models for 2D
van der Waals heterojunctions with a smaller lattice mismatch for
different 2D materials. Its basic idea is based on the rotation
matrix of extended orthogonal lattices for different 2D materials to
search their minimum matching supercell models in 2D van der Waals
heterojunctions.


\section{Computational methods}

Our first-principles calculations are based on the density
functional theory (DFT) implemented in the VASP
package.\cite{PRB_47_558_1993_VASP} All the geometry structures are
fully relaxed until energy and forces are converged to 10$^{-5}$
$eV$ and 0.01 $eV$/{\AA}, respectively. Dipole correction is
employed to cancel the errors of electrostatic potential, atomic
forces and total energy, caused by periodic boundary
condition~\cite{PRB_51_4014_1995_DipoleCorrection}. Very fine points
in the surface Brillouin zone sampled with a regular mesh are used
for calculating the tiny band gaps at the Dirac points of graphene
and silicene in our recently studied 2D van der Waals
heterojunctions.

Because conventional exchange-correlation functionals in the DFT
calculations, such as the generalized gradient approximation of
Perdew, Burke, and Ernzerhof (GGA-PBE)~\cite{PRL_77_3865_1996}, fail
to adequately describe the long-range weak van der Waals
interactions, different methods, such as semi-empirical long-range
dispersion correction (DFT-D) proposed by Grimme $et$
$al.$~\cite{JCC_25_1463_2004_DFT_D1, JCC_27_1787_2006_DFT_D2,
JCC_32_1456_2011_DFT_D3} and non-empirical van der Waals density
functional (vdW-DF) scheme proposed by Dion $et$
$al.$~\cite{PRL_92_246401_2004, PRL_103_096102_2009,
PRB_82_081101_2010}, have been developed. Table~\ref{vdW} shows
different methods implemented in VASP calculated equilibrium
interfacial distance and binding energy per carbon atom for bilayer
graphene ($a$ = $b$ = 2.47 {\AA}) compared with experimental
measurements~\cite{PRB_69_155406_2004, PRB_85_205402_2012}. In our
recent theoretical studies on 2D van der Waals heterojunctions, we
choose the semi-empirical DFT-D2
method~\cite{JCC_27_1787_2006_DFT_D2} due to its widely used and
good description of long-range weak van der Waals interactions in
molecular surface adsorption and layered structure
systems~\cite{JPCC_111_11199_2007, PCCP_10_2722_2008,
NanoLett_11_5274_2011, PRB_83_245429_2011, PRB_85_125415_2012,
PRB_85_235448_2012, JPCL_4_2158_2013, Nanoscale_5_9062_2013,
PCCP_15_5753_2013, PCCP_16_6957_2014, PCCP_16_22495_2014}.

\begin{table}
\caption{Different methods implemented in VASP calculated
equilibrium interfacial distance $D_0$ ({\AA}) and binding energy
per carbon atom $E_b$ ($meV$) for bilayer graphene compared with
experimental measurements.}
\begin{tabular}{ccccc} \\ \hline \hline
Bilayer graphene &  $D_0$  &  $E_b$  \ \\
\hline
GGA (PBE)        &  3.90  &  -0.04  \ \\
DFT-D2 (PBE)     &  3.25  &  -25.2  \ \\
vdW-DF (PBE)     &  3.47  &  -33.6  \ \\
vdW-DF (optB88)  &  3.32  &  -32.0  \ \\
Exp.~\cite{PRB_69_155406_2004, PRB_85_205402_2012}      &  3.35  &  -26.0  \ \\
\hline \hline
\end{tabular}
\label{vdW}
\end{table}

The charge carrier (hole and electron) concentration of doping
graphene in 2D van der Waals heterojunctions can be estimated by the
linear dispersion around the Dirac point of
graphene\cite{RMP_81_109_2009}
\begin{equation}
N_{h/e}=\frac{(\bigtriangleup{E_D})^2}{\pi(\hbar\nu_{F})^{2}}
\end{equation}
where $\bigtriangleup$$E_D$ is the shift of graphene's Dirac point
($E_D$) relative to the Fermi level ($E_F$), that is
$\bigtriangleup$$E_D$ = $E_D$ - $E_F$.

To study the optical properties of 2D van der Waals heterojunctions,
the frequency-dependent dielectric matrix is
calculated~\cite{PRB_73_045112_2006}. The imaginary part of
dielectric matrix is determined by a summation over states as
\begin{equation}
\begin{aligned}
\varepsilon_{\alpha\beta}^{\prime\prime}=&\frac{4\pi^2e^2}{\Omega}\mathop{\lim}\limits_{q \to 0}\frac{1}{q^2}\sum_{c,v,\textbf{k}}2w_{\textbf{k}}\delta(\epsilon_{c\textbf{k}}-\epsilon_{v\textbf{k}}-\omega) \\
                               &\times\langle\mu_{c\textbf{k}+\textbf{e}_{_{\alpha}}q}|\mu_{v\textbf{k}}\rangle\langle\mu_{c\textbf{k}+\textbf{e}_{_{\beta}}q}|\mu_{v\textbf{k}}\rangle^{*}
\end{aligned}
\end{equation}
where, $\Omega$ is the volume of the primitive cell,
$w_{\textbf{k}}$ is the $\textbf{k}$ point weight, $c$ and $v$ are
the conduction and valence band states respectively,
$\epsilon_{c\textbf{k}}$ and $\mu_{c\textbf{k}}$ are the eigenvalues
and wavefunctions at the $\textbf{k}$ point respectively, and
$\textbf{e}_{_{\alpha}}$ are the unit vectors for the three
Cartesian directions. In order to accurately calculate the optical
properties of 2D van der Waals heterojunctions, a large four times
regular mesh for the surface Brillouin zone, a large number of empty
conduction band states (two times more than the number of valence
band) and frequency grid points are adopted in optical properties
calculations compared with conventional electronic structure
calculations.


\section{Van der Waals heterojunctions}
\begin{figure*}[htb]
\centering
\includegraphics[width=0.5\textwidth]{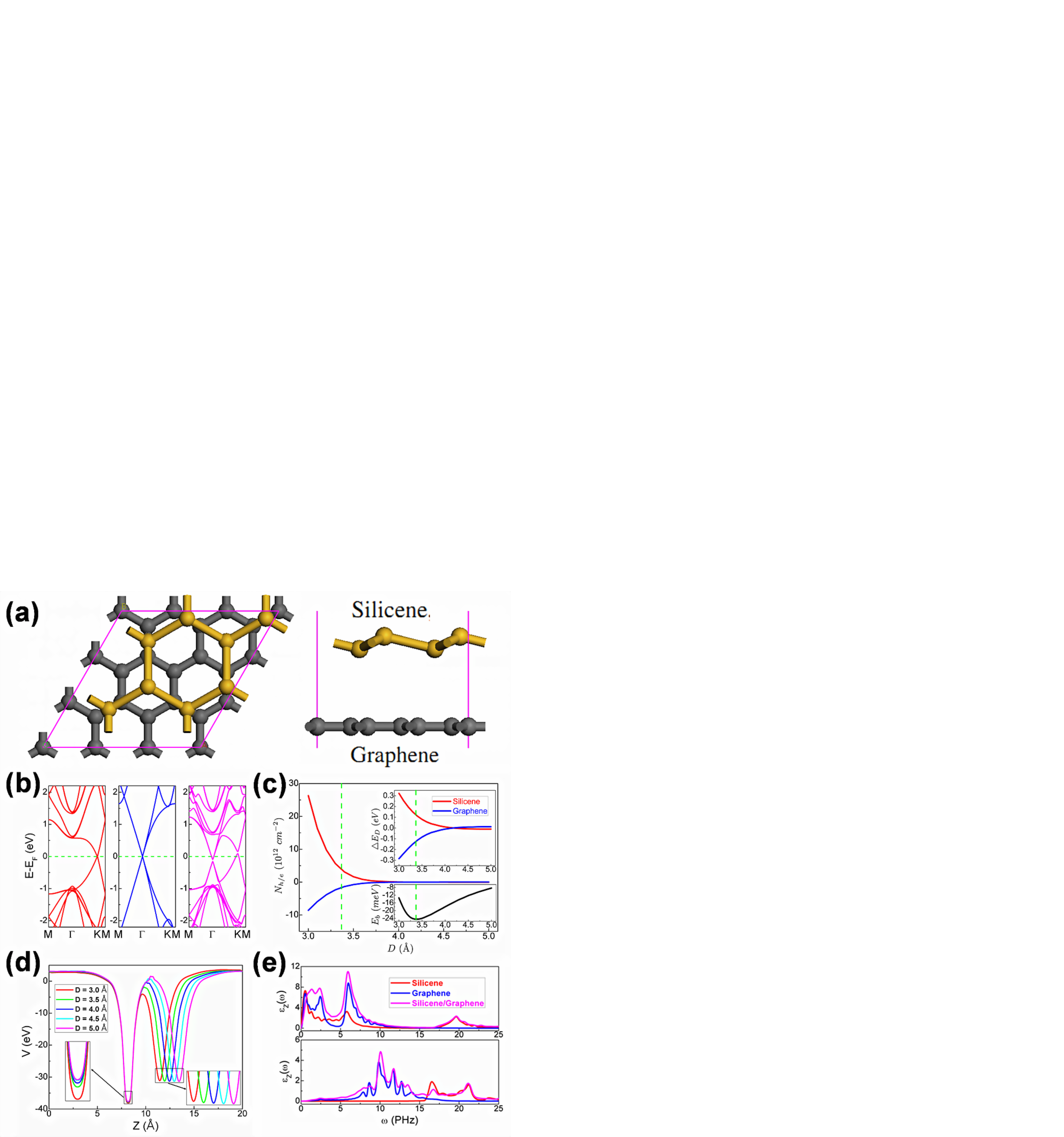}
\caption{(Color online) (a) Geometric structures of G/S
heterojunction (top and side views). The gray and yellow balls
denote carbon and silicon atoms, respectively. (b) Electronic band
structures of silicene, graphene and G/S heterojunction. The Fermi
level is set to zero and marked by green dotted lines. (c) The
doping charge carrier concentrations $N_{h/e}$ ($10^{12}$ $cm^{-2}$)
of silicene and graphene of in G/S heterojunction as a function of
interfacial distance $D$ ({\AA}). The binding energies $E_{b}$
($meV$) per atom and the Dirac points $\bigtriangleup$$E_D$ ($eV$)
of silicene and graphene shift relative to the Fermi level are shown
in the inset. The equilibrium spacing is denoted by green dotted
lines. (d) XY-averaged electrostatic potentials of G/S
heterojunction at different interfacial distances $D$ ({\AA}) in the
Z direction. Depths of potential wells of graphene and silicene are
shown in the inset. (e) Imaginary part ($\epsilon^{\prime\prime}$)
of frequency ($E=\hbar\omega$) dependent dielectric function
(parallel and perpendicular) for pristine silicene and graphene
monolayers as well as corresponding G/S heterojunction. (Reproduced
from Ref.~\cite{JCP_139_154704_2013}, with permission from American
Institute of Physics.)} \label{fig:GS}
\end{figure*}
\begin{figure*}[htb]
\centering
\includegraphics[width=0.5\textwidth]{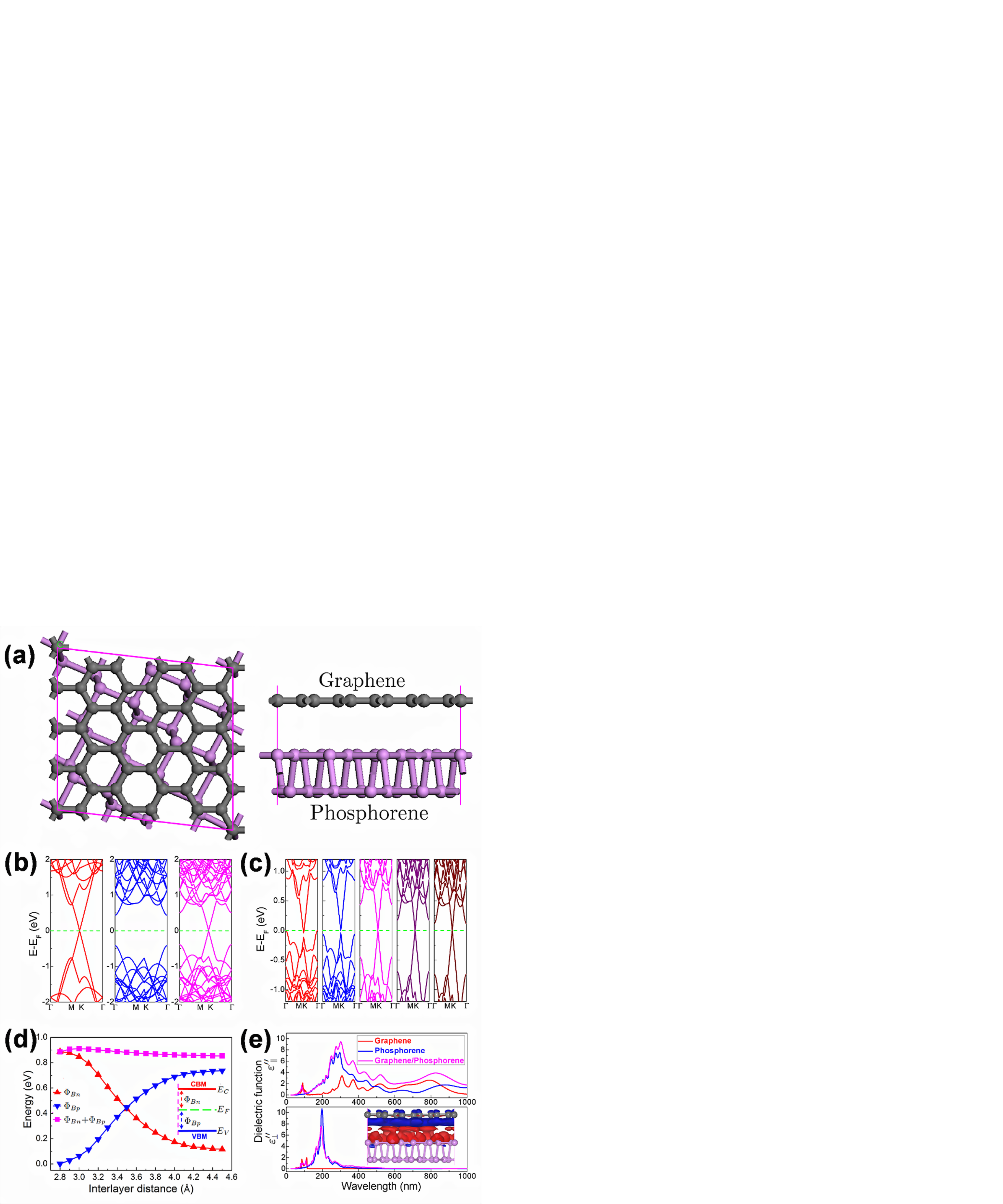}
\caption{(Color online) (a) Geometric structures of G/P
heterojunction (top and side views). The gray and violet balls
denote carbon and phosphorus atoms, respectively. (b) Electronic
band structures of graphene, phosphorene and G/P heterojunction. The
Fermi level is set to zero and marked by green dotted lines. (c)
Electronic band structures of G/P heterojunction at different
interfacial distances $D$ = 2.8, 3.0, 3.5, 4.0 and 4.5 {\AA}. (d)
Schottky barriers $\Phi$$_{Bn}$, $\Phi$$_{Bp}$ and
$\Phi$$_{Bn}$+$\Phi$$_{Bp}$ in G/P heterojunction as a function of
interfacial distance. (e) Imaginary part of frequency dependent
dielectric function (parallel and perpendicular) for pristine
graphene and phosphorene monolayers as well as corresponding G/P
heterojunction. Differential charge density (0.002 $e$/{\AA}$^3$) of
G/P heterojunction is shown in the insert. The red and blue regions
indicate electron increase and decrease, respectively. (Reproduced
from Ref.~\cite{JMCC_3_4756_2015}, with permission from Royal
Society of Chemistry.)} \label{fig:GP}
\end{figure*}
\begin{figure*}[htb]
\centering
\includegraphics[width=0.5\textwidth]{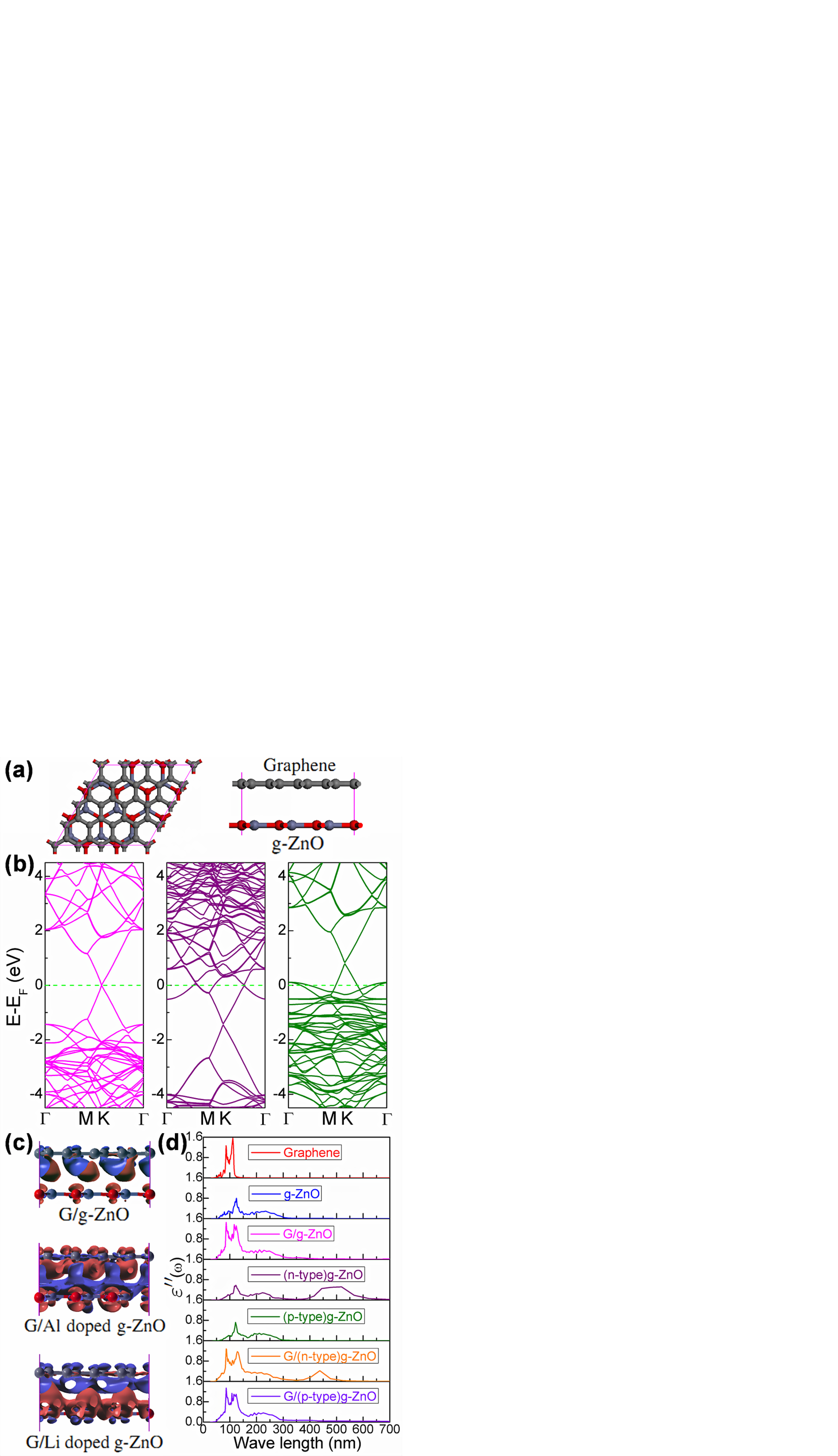}
\caption{(Color online) (a) Geometric structures structures of
G/g-ZnO heterojunction (top and side views). The gray, red and blue
balls denote carbon, oxygen and zinc atoms, respectively. (b)
Electronic band structures of G/g-ZnO, G/Al doped g-ZnO, and G/Li
doped g-ZnO heterojunctions. The Fermi level is set to zero and
marked by red dotted lines. (c) Differential charge density (0.002
$e$/{\AA}$^3$) for G/g-ZnO, G/Al doped g-ZnO, and G/Li doped g-ZnO
heterojunctions. The red and blue regions indicate electron increase
and decrease, respectively. (d) Imaginary part
($\epsilon^{\prime\prime}$) of dielectric function (perpendicular)
for graphene, g-ZnO monolayers and G/g-ZnO heterojunctions.
(Reproduced from Ref.~\cite{JCP_138_124706_2013}, with permission
from American Institute of Physics.)} \label{fig:GZnO}
\end{figure*}
\begin{figure*}[htb]
\centering
\includegraphics[width=0.5\textwidth]{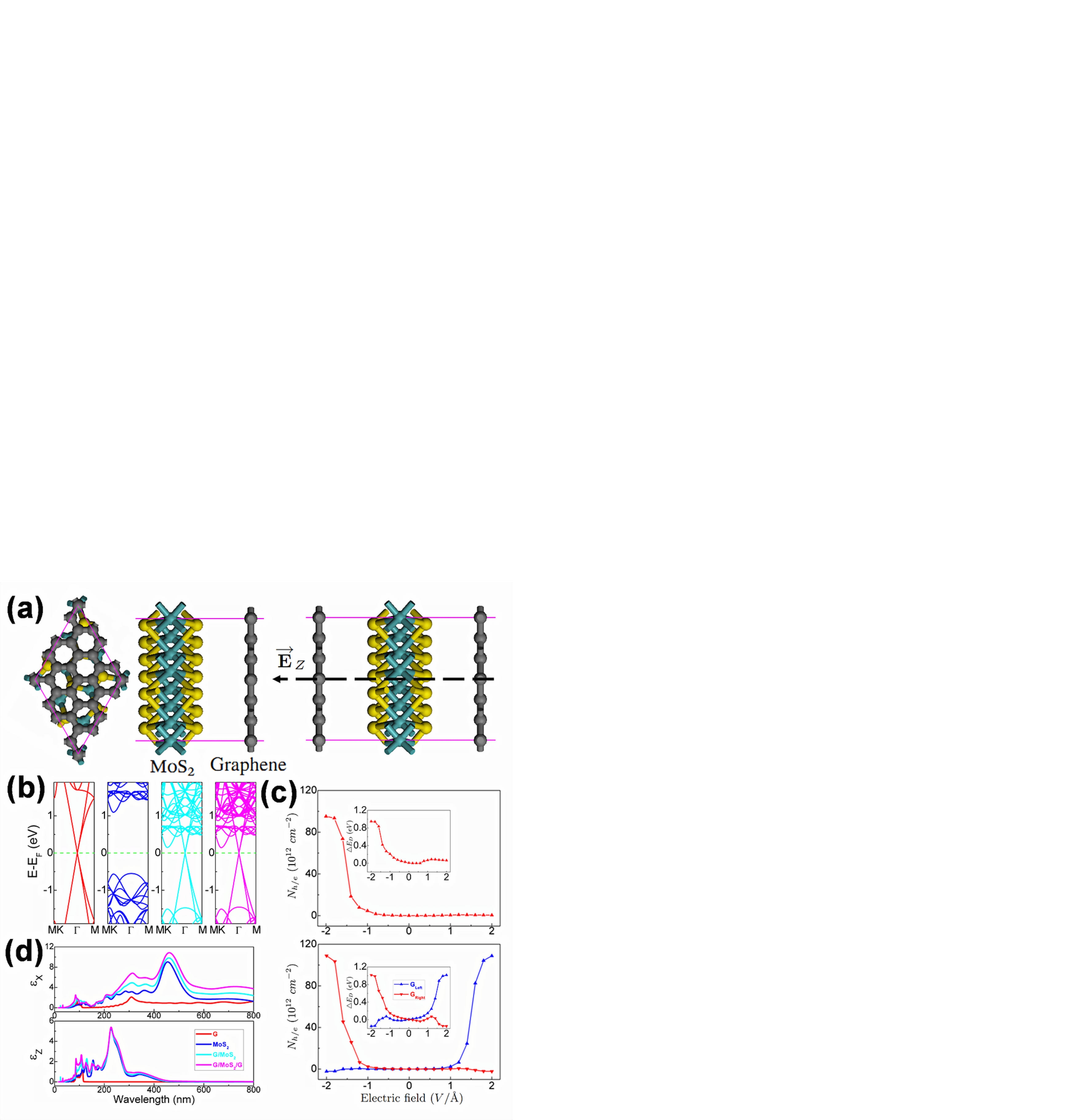}
\caption{(Color online) (a) Geometric structures of G/MoS$_2$ and
G/MoS$_2$/G heterojunctions. The gray, yellow and blue balls denote
carbon, sulfur and molybdenum atoms, respectively. (b) Electronic
band structures of graphene and MoS$_2$ as well as corresponding
G/MoS$_2$ and G/MoS$_2$/G heterojunctions. The Fermi level is set to
zero and marked by green dotted lines. (c) The doping charge carrier
concentrations $N_{h/e}$ ($10^{12}$ $cm^{-2}$) of graphene in
G/MoS$_2$ and G/MoS$_2$/G heterojunctions as a function of vertical
electric field $E$ ($V$/{\AA}). The energy shift
$\bigtriangleup$$E_D$ ($eV$) of graphene's Dirac point relative to
the Fermi level is shown in the inset. (d) Imaginary part of
dielectric function ($\varepsilon^{\prime\prime}$) of pristine
graphene MoS$_2$ monolayers as well as corresponding G/MoS$_2$ and
G/MoS$_2$/G heterojunctions. (Reproduced from
Ref.~\cite{arXiv:1411.0357_2014})} \label{fig:GMoS2}
\end{figure*}
\begin{figure*}[htb]
\centering
\includegraphics[width=0.5\textwidth]{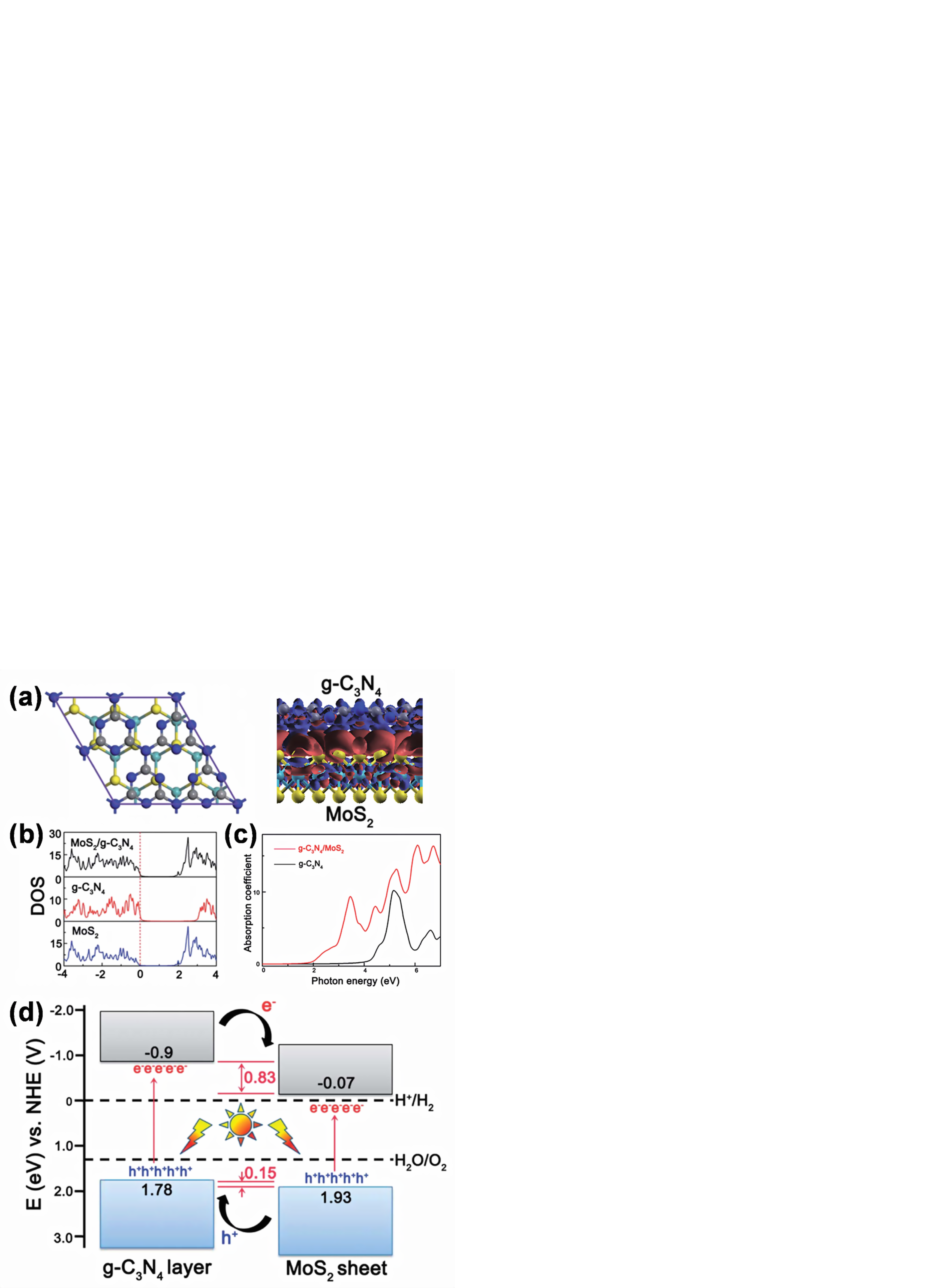}
\caption{(Color online) (a) Geometric structures and differential
charge density of g-C$_3$N$_4$/MoS$_2$ heterojunction. The gray,
blue, yellow and blue balls denote carbon, nitrogen, sulfur and
molybdenum atoms, respectively. The red and blue regions indicate
electron increase and decrease, respectively. (b) Density of states
of g-C$_3$N$_4$/MoS$_2$ heterojunction, g-C$_3$N$_4$ and MoS$_2$
monolayers. The Fermi level is set to zero and marked by green
dotted lines. (c) Absorption coefficients of g-C$_3$N$_4$ monolayer
and g-C$_3$N$_4$/MoS$_2$ heterojunction. (d) Schematic illustration
of carrier transfer and separation in g-C$_3$N$_4$/MoS$_2$
heterojunction. (Reproduced from Ref.~\cite{JPCA_2_7960_2014}, with
permission from American Chemical Society.)} \label{fig:C3N4MoS2}
\end{figure*}

We first check some common properties in our recently studied 2D van
der Waals heterojunctions, including the Graphene/Silicene
(G/S)~\cite{JCP_139_154704_2013}, Graphene/Phosphorene
(G/P)~\cite{JMCC_3_4756_2015}, Graphene/g-ZnO
(G/g-ZnO)~\cite{JCP_138_124706_2013}, Graphene/MoS$_2$
(G/MoS$_2$)~\cite{arXiv:1411.0357_2014} and
g-C$_3$N$_4$/MoS$_2$~\cite{JPCA_2_7960_2014} heterojunctions as
shown in
Fig.~\ref{fig:GS},~\ref{fig:GP},~\ref{fig:GZnO},~\ref{fig:GMoS2}
and~\ref{fig:C3N4MoS2}, respectively. We find that the weak van der
Waals interlayer interactions in 2D van der Waals heterojunctions
can induce new properties and phenomena, such as bandgap opening,
charge transfer and new optical absorption.

Firstly, we find that weak van der Waals interactions domain at the
interfaces of these 2D van der Waals heterojunctions.
Table~\ref{DFT-D2} shows our recently studied 2D van der Waals
heterojunctions all have typical vdW equilibrium spacings with
corresponding small binding energies, thus, they can be used as
ideal substrates for each other with their intrinsic electronic
structures undisturbed in 2D van der Waals heterojunctions.
Furthermore, we confirm that the small lattice mismatch of about 2\%
for special supercells of 2D materials obtained by our LatticeMatch
program has little effect on their electronic properties in these 2D
van der Waals heterojunctions as shown their electronic band
structures and density of states in
Fig.~\ref{fig:GS}b,~\ref{fig:GP}b,~\ref{fig:GZnO}b,~\ref{fig:GMoS2}b
and~\ref{fig:C3N4MoS2}b.

\begin{table}
\caption{DFT-D2 calculated equilibrium interfacial distance $D_0$
({\AA}) and binding energy per carbon atom $E_b$ ($meV$) in our
studied 2D van der Waals heterojunctions}
\begin{tabular}{ccccc} \\ \hline \hline
DFT-D2                 &  $D_0$  &  $E_b$  \ \\
\hline
Graphene/Silicene      &  3.37  &  -24.7  \ \\
Graphene/Phosphorene   &  3.43  &  -24.7  \ \\
Graphene/g-ZnO         &  3.14  &  -51.0  \ \\
Graphene/MoS$_2$       &  3.37  &  -20.5 \ \\
g-C$_3$N$_4$/MoS$_2$   &  2.97  &  -22.4  \ \\
\hline \hline
\end{tabular}
\label{DFT-D2}
\end{table}

Secondly, small band gaps are opened at the Dirac points of graphene
in graphene based 2D van der Waals heterojunctions due to weak van
der Waals interactions at the interfaces, though its linear
Dirac-like dispersion relation around the Fermi level is still
preserved. Notice that induced graphene band gaps in 2D van der
Waals heterojunctions are typically sensitive to other external
conditions, such as interlayer separation~\cite{APL_98_083103_2011},
showing that the band gap values increase gradually with the
interlayer separation decrease, thus tunable, with potential
applications for graphene-based field effect transistors.

Thirdly, charge redistribution and transfer always occur at the
interfaces in 2D van der Waals heterojunctions due to the interlayer
coupling. In our recently studied 2D van der Waals heterojunctions,
the G/S heterojunction belongs to the Metal/Metal type, the G/g-ZnO,
G/P, and G/MoS$_2$ heterojunctions belong to the Metal/Semiconductor
type, the g-C$_3$N$_4$/MoS$_2$ heterojunction belongs to the
Semiconductor/Semiconductor type. For metal, they have different
work functions (The energy difference between the vacuum level and
the Fermi level). For semiconductors, they have different
ionizations (The energy difference between the vacuum level and the
conduction band minimum) and nucleophilics (The energy difference
between the vacuum level and the valence band maximum) potentials.
Based on the Schottky-Mott model~\cite{PR_71_717_1947}, different
work functions, ionizations and nucleophilics can induce charge
transfer as well as well-separated electron-hole pairs at the
interfaces in 2D van der Waals heterojunctions (Fig.~\ref{fig:GS}c,
~\ref{fig:GP}e, ~\ref{fig:GZnO}c and ~\ref{fig:C3N4MoS2}a).
Furtheremore, inhomogeneous surfaces also can results in charge
redistribution and form intralayer electron-hole puddles at the
interfaces in 2D van der Waals heterojunctions as shown in
Fig.~\ref{fig:GZnO}b. These new properties of charge redistribution
and transfer at the interfaces can significantly enhance the
conductivity and generate new catalytic activities in 2D van der
Waals heterojunctions.

Fourthly, interlayer interaction~\cite{JPCL_3_3330_2012} and charge
transfer~\cite{JACS_134_4393_2012} in 2D van der Waals
heterojunctions may induce new optical transitions, though pristine
monolayers themselves display outstanding optical properties. We
find that 2D van der Waals heterojunctions always exhibit wider
absorption range and stronger optical absorption compared with their
single monolayers as shown in
Fig.~\ref{fig:GS}e,~\ref{fig:GP}e,~\ref{fig:GZnO}d,~\ref{fig:GMoS2}d
and~\ref{fig:C3N4MoS2}c, because electrons can now be directly
excited between different layers at the interfaces of
heterojunctions. These new optical transitions in 2D van der Waals
heterojunctions are expected to be with a great potential in
photocatalytic and photovoltaic devices.

In the following of this review, we primarily focus on some special
features of our recently studied 2D van der Waals heterojunctions
that distinguish them from other previously studied 2D van der Waals
heterojunctions.

\subsection{Graphene/Silicene}

Graphene and silicene are metallic but they have different work
functions. Therefore, both silicene and graphene can be doped in
hybrid G/S heterojunction as plotted in Fig.~\ref{fig:GS}b, and
their Dirac points respectively shift 0.12 $eV$ above and below the
Fermi levels in hybrid G/S heterojunction, inducing weak p-type and
n-type doping of silicene and graphene, respectively. Our calculated
charge carrier concentrations are $N_{h}$(S) = 3.8$\times$$10^{12}$
$cm^{-2}$ and $N_{e}$(G) = 1.6$\times$$10^{12}$ $cm^{-2}$ for
silicene and graphene in hybrid S/G heterojunction, respectively.
Notice that these charge carrier concentrations are more than 2
orders of magnitude larger than the intrinsic charge carrier
concentration of graphene at room temperature ($n$ =
$\pi$$k_{B}^2$$T^2$/6$\hbar$$\nu_{F}^2$ = 6$\times$$10^{10}$
$cm^{-2}$). Furthermore, we find that the charge carrier
concentrations of both silicene and graphene in hybrid G/S
heterojunction can be tuned via the interfacial spacing as shown in
Fig.~\ref{fig:GS}c. Their charge carrier concentrations decrease
gradually with the interlayer separation increase and a conversion
of doping types of silicene and graphene occurs when their
interfacial distance artificially increases to above 4.2 {\AA}.
Therefore, the self-doping phenomenon in hybrid G/S heterojunction
provides a effective and tunable way for new optoelectronic devices.

Here, we reveal the origin intrinsic mechanism of tunable
self-doping in hybrid G/S heterojunction. We observe that
$\bigtriangleup$$E_D$ is already unchanged for both silicene and
graphene as the interfacial distance increases larger than 4.6
{\AA}, but $\bigtriangleup$$E_D$ converges to different values
($\bigtriangleup$$E_D$$(S)$ = -0.01 $eV$ and
$\bigtriangleup$$E_D$$(G)$ = 0.01 $eV$). It can be interpreted by
their different work functions, $W_F$$(S)$ = 4.6 $eV$ and $W_F$$(G)$
= 4.4 $eV$. Based on the Schottky-Mott model~\cite{PR_71_717_1947},
few electrons should be transferred from graphene to silicene,
resulting in weak n-type and p-type doping of silicene and graphene,
respectively. When the interfacial distance decreases, weak overlap
of electronic states between silicene and graphene is enhanced,
which increase the charge transfer and electric double layer at the
G/S interface, shifting up and down the energy levels of silicene
and graphene respectively as shown in Fig.~\ref{fig:GS}d.
Furthermore, the tunneling energy barrier for electrons at the G/S
interface is also reduced by their interface interactions.
Therefore, when silicene and graphene are close to each other
(smaller than 4.2 {\AA}), few electrons are transferred from
silicene to graphene, resulting in weak p-type and n-type doping of
silicene and graphene, respectively. That is why a reversion of
doping types of silicene and graphene occurs as the interfacial
distance is 4.2 {\AA}. In previous studies, similar feature of
tunable doping types and charge carrier concentration of graphene on
metal substrates~\cite{PRL_101_026803_2008} have been also reported
experimentally and theoretically.

\subsection{Graphene/Phosphorene}

Monolayer phosphorene is semiconducting with a direct band gap of
0.85 $eV$ (Fig.~\ref{fig:GP}b), which is different form graphene.
Therefore, a Schottky contact can be formed between metallic
graphene and semiconducting phosphorene as shown in
Fig.~\ref{fig:GP}d. Based on the Schottky-Mott
model~\cite{PR_71_717_1947} at the metal/semiconductor
interface~\cite{NanoLett_13_509_2013}, a n-type Schottky barrier
($\Phi$$_{Bn}$) is defined as the energy difference between the
Fermi level ($E$$_{F}$) and the conduction band minimum ($E$$_{C}$),
that is $\Phi$$_{Bn}$ = $E$$_{C}$ - $E$$_{F}$. Similarly, a p-type
Schottky barrier ($\Phi$$_{Bp}$) is defined as the energy difference
between the Fermi level ($E$$_{F}$) and valence band maximum
($E$$_{V}$), that is $\Phi$$_{Bp}$ = $E$$_{F}$ - $E$$_{V}$.
Furthermore, the sum of two types (n- and p-type) of Schottky
barrier is approximately equal to the band gap value ($E_G$) of
semiconductor, that is $\Phi$$_{Bn}$ + $\Phi$$_{Bp}$ $\approx$
$E_G$. As shown in Fig.~\ref{fig:GP}c, the Dirac point of graphene
moves from the conduction band to the valence band of phosphorene as
the interfacial distance decreases from 4.5 to 2.8 {\AA}, inducing a
transition from n-type Schottky contact to p-type Schottky contact
at the interface. The conversion of Schottky contact type in G/P
heterojunction occurs when their interfacial distance increases to
below 3.5 {\AA}.

We find that when the interfacial distance increases larger than 4.5
{\AA}, the Dirac point of graphene is close to the conduction band
of phosphorene, forming a n-type Schottky contact with a small
barrier $\Phi$$_{Bn}$ = 0.12 $eV$ at the interface as shown in
Fig.~\ref{fig:GP}d. That is because the work function of graphene
(4.3 $eV$)~\cite{JCP_139_154704_2013} is close to the nucleophilic
potential of phosphorene (4.1 $eV$). When the interfacial distance
decreases from 4.5 to 3.0 {\AA}, the charge transfer and chemical
interactions between graphene and phosphorene are
enhanced~\cite{NanoLett_13_509_2013}. As the interfacial distance
artificially decreases to 3.0 {\AA}, the Dirac point of graphene is
close to the valence band of phosphorene, forming a p-type Schottky
contact with a negligible p-type Schottky barrier of $\Phi$$_{Bp}$ =
0.06 $eV$ at the interface as shown in Fig.~\ref{fig:GP}d.
Furthermore, strong interlayer-interactions even can induce weak
n-type doping in graphene when the interfacial distance artificially
decreases to 2.8 {\AA} as shown in Fig.~\ref{fig:GP}c. Therefore,
G/P heterojunction has tunable Schottky contacts and barriers as
varying the the interfacial distance.

Notice that the Schottky contact in G/P heterojunction is very
different to traditional metal-semiconductor Schottky ones in two
important ways. One is that graphene is adsorbed physically on
phosphorene in ultrathin van der Waals heterojunction without
dangling bonds at the interface. Another is the Schottky contacts
and barriers can be adjusted sensitively by varying the interfacial
distance. Furthermore, atom doping in graphene and applying vertical
electric fields can also be used to modify the work function of
graphene and then to adjust the Schottky contacts and barriers in
G/P heterojunction. Therefore, G/P heterojunction can be used for
tunable Schottky diodes in nanoelectronics.

\subsection{Graphene/g-ZnO}

Monolayer g-ZnO is a wide-gap semiconductor. When forming a hybrid
G/undoped g-ZnO heterojunction, the Fermi level still remains in the
induced gap, showing that little charge transfer between graphene
and undoped g-ZnO monolayer. This is also supported by the
differential charge density as plotted in Fig.~\ref{fig:GZnO}c.
However, the inhomogeneous g-ZnO substrate can induce special charge
redistribution in the graphene plane, forming intralayer
electron-hole puddles as plotted in Fig.~\ref{fig:GZnO}c. This
effect can significantly enhance the electron conductivity and
generate new photovoltaic activities~\cite{JACS_134_4393_2012}.

For hybrid G/doped g-ZnO heterojunctions, we observe that different
charge rearrangement occurs in the heterojunctions as shown in
Fig.~\ref{fig:GZnO}c. We find that Al and Li doped g-ZnO monolayers
have a work function either 1.2 $eV$ smaller or the same amount
larger than that of graphene. Based on the Schottky-Mott
model\cite{PR_71_717_1947}, the Dirac points of hybrid G/doped g-ZnO
heterojunctions are moved to below and above the Fermi level,
respectively. As shown in Fig.~\ref{fig:GZnO}c, electron-hole pairs
are well separated at the interfaces of hybrid G/doped g-ZnO
heterojunctions. The calculated charge carrier density is about
$N_{e}$(G/n-type g-ZnO) = 3.5$\times$$10^{13}$ $cm^{-2}$ and
$N_{h}$(G/p-type g-ZnO) = 1.1$\times$$10^{13}$ $cm^{-2}$. Notice
that these charge carrier concentrations are more than 3 orders of
magnitude larger than the intrinsic charge carrier concentration of
graphene at room temperature ($n$ =
$\pi$$k_{B}^2$$T^2$/6$\hbar$$\nu_{F}^2$ = 6$\times$$10^{10}$
$cm^{-2}$). Therefore, the charge transfer in hybrid G/doped g-ZnO
heterojunctions can induce effective and tunable electron or hole
doping in graphene for graphene-based optoelectronic devices, such
as Schottky diodes and p-n junctions.

\subsection{Graphene/MoS$_2$}

High-performance field-effect tunneling transistors have been
achieved experimentally~\cite{Science_340_1311_2013,
ACSNano_7_7021_2013} in hybrid graphene and MoS$_2$ heterojunctions.
Thus, the electronic properties of hybrid G/MoS$_2$ and G/MoS$_2$/G
heterojunctions affected by applying vertical electric fields are
very desirable as shown in Fig.~\ref{fig:GMoS2}c. We find taht
negative vertical electric fields can induce p-type doping of
graphene in hybrid G/MoS$_2$ heterojunctions. But, positive electric
fields almost have on effect on the electronic properties of hybrid
G/MoS$_2$ heterojunctions. This is because electrons can easily from
the Dirac point of graphene to the conduction band of MoS$_2$ but
difficulty from the valence band of MoS$_2$ to the Dirac point of
graphene due to the work function (4.3
$eV$)~\cite{JCP_138_124706_2013} of graphene close to the electronic
affinity (4.2 $eV$) of monolayer MoS$_2$.

We find that vertical electric fields can generate strong p-type but
weak n-type doping of graphene at both negative and positive
electric fields due to the symmetry in hybrid G/MoS$_2$/G
heterojunctions. The calculated charge carrier concentrations in
hybrid G/MoS$_2$ and G/MoS$_2$/G heterojunctions are shown in
Fig.~\ref{fig:GMoS2}c. These values are more than 3 orders of
magnitude larger than the intrinsic charge carrier concentration of
graphene at room temperature ($n$ =
$\pi$$k_{B}^2$$T^2$/6$\hbar$$\nu_{F}^2$ = 6$\times$$10^{10}$
$cm^{-2}$). The doping charge carrier concentrations of graphene in
hybrid G/MoS$_2$ and G/MoS$_2$/G heterojunctions are increased with
the vertical electric fields. Furthermore, the electron-hole pairs
are well separated in hybrid G/MoS$_2$/G sandwiched heterojunctions
with more excellent applications\cite{Science_340_1311_2013,
ACSNano_7_7021_2013} compared with hybrid G/MoS$_2$ heterojunctions.
Therefore, the field-effect in hybrid graphene and MoS$_2$
heterojunctions is effective and tunable for high-performance FETs
and p-n junctions.

\subsection{g-C$_3$N$_4$/MoS$_2$}

Monolayer g-C$_3$N$_4$ and MoS$_2$ are all semiconductors with
direct band gaps of 2.7 and 2.0 $eV$, respectively. g-C$_3$N$_4$ can
be used to achieve both half reactions of water splitting under
visible light, meaning that its bandgap cover both the water
reduction and water oxidation potentials. However, the
photocatalytic H$_2$ production activity of pristine g-C$_3$N$_4$
remains poor, strongly depending on the surface co-catalysts. On the
other hand, MoS$_2$ has been extensively investigated as a promising
electrocatalyst for H$_2$ evolution due to its high abundance and
low cost. According to some experimental studies, MoS$_2$ can serve
as a suitable co-catalysts to composite with g-C$_3$N$_4$ to show
enhanced photocatalytic performance.

Our calculated band alignment between the g-C$_3$N$_4$ and MoS$_2$
monolayers reveals that the conduction band minimum and valence band
maximum of the g-C$_3$N$_4$ monolayer are higher by about 0.83 $eV$
and 0.15 $eV$ respectively than those of the MoS$_2$ monolayer. This
predicted type-II band alignment ensures the photogenerated
electrons easily migrate from the g-C$_3$N$_4$ monolayer to the
MoS$_2$ monolayer, and leads to the high hydrogen-evolution reaction
activity. The charge transfer between g-C$_3$N$_4$ and MoS$_2$
results in a polarized field within the interface region, which will
benefit the separation of photogenerated carriers. The calculated
optical absorption curves verify that this proposed layered
nanocomposite is a good light-harvesting semiconductor. These
findings indicate that MoS$_2$ is a promising candidate as a
non-noble metal co-catalyst for g-C$_3$N$_4$ photocatalysts, and
also provide useful information for understanding the observed
enhanced photocatalytic mechanism in experiments.


\section{Summary and conclusions}

In the present work, we review our recent theoretical studies on the
structural, electronic, optical and electrical properties of 2D van
der Waals heterojunctions via density functional theory
calculations, including the Graphene/Silicene, Graphene/Phosphorene,
Graphene/g-ZnO, Graphene/MoS$_2$ and g-C$_3$N$_4$/MoS$_2$
heterojunctions. We find that the weak interlayer van der Waals
interactions in these 2D heterojunctions can induce new properties
and phenomena, such as bandgap opening, charge transfer and new
optical absorption. Furthermore, our recently studied 2D van der
Waals heterojunctions have some special features which distinguish
them from other previously studied 2D van der Waals heterojunctions.
With excellent structural, electronic, electrical and optical
properties combined, 2D van der Waals heterojunctions are expected
to be with great applications in efficient electronic,
electrochemical, photovoltaic, photoresponsive and memory devices.


\section{Acknowledgements}

This work is partially supported by the National Key Basic Research
Program (2011CB921404), by NSFC (11404109, 21121003, 91021004,
21233007, 21222304), by CAS (XDB01020300). This work is also
partially supported by the Scientific Discovery through Advanced
Computing (SciDAC) program funded by U.S. Department of Energy,
Office of Science, Advanced Scientific Computing Research and Basic
Energy Sciences (W. H.). We thank the National Energy Research
Scientific Computing (NERSC) center, and the USTCSCC, SC-CAS,
Tianjin, and Shanghai Supercomputer Centers for the computational
resources.

\section{Appendix A. Supplementary material}

Supplementary material associated with this review article can be
found, in the online version, at http://dx.doi.org/10.XXXX/XX.XX.



\end{document}